# О «фазовых переходах» в магнитных мезоскопических системах

*О.П. Мартыненко, В. В. Махро, И.Г. Махро*

*Обсуждаются особенности перехода между классическим и квантовым режимами в бистабильных высокоспиновых системах, описываемых гамильтонианом* $H = -DS_z^2 - H_x S_x$.

Вполне понятен интерес, проявляемый в последнее время к макро- и мезоскопическим квантовым системам. Их исследование важно, прежде всего, с точки зрения проверки основ квантовой механики [2], и, в то же время, представляет значительный прикладной интерес. Очень перспективными в этом отношении являются эффекты туннелирования в магнитных мезоскопических системах [3, 4, 5, 6, 7]. Объектом пристального внимания и экспериментаторов, и теоретиков недавно стали молекулярные магниты, в особенности, система $Mn_{12}Ac$ [8, 9, 10]. В молекуле этого соединения 12 ионов Mn сильно ферромагнитно связаны посредством суперобмена через кислородные мостики, что позволяет рассматривать такую молекулу, как магнитный кластер со спином 10 [9]. В таком кластере состояния с противоположной ориентацией спинов разделены энергетическим барьером с эффективной высотой $|D|S^2$, составляющей 72 – 86 $k_B$. Переходы между двумя состояниями могут возникать либо под действием внешнего поля, либо в результате термоактивации, либо, наконец, как следствие туннельного эффекта. Здесь мы ограничимся рассмотрением последних двух механизмов.

Будем исходить из гамильтониана

$$H = -DS_z^2 - H_x S_x \quad , \quad (1)$$

где $S \gg 1$. В настоящее время считается, что эта модель достаточно адекватно описывает ситуацию в $Mn_{12}Ac$ [6]. Сравнение термоактивационного и туннельного режимов удобно проводить после отображения спиновой задачи на задачу о движении частицы [11]. Гамильтониан для частицы, соответствующий спиновому гамильтониану (1), в этом случае будет

$$H = -\frac{\nabla^2}{2m} + U(x) \quad , \quad (2)$$

где

$$U(x) = (S + 1/2)^2 D(h_x^2 \sinh^2 x - 2h_x \cosh x), \quad (3)$$

причём $m \equiv 1/2D$ и $h_x \equiv \dfrac{H_x}{(2S+1)D}$.

Подчеркнем, что при всех значениях поперечного поля $0 < h_x < 1$ функция (3) описывает гладкий двухъямный потенциал (рис. 1). Таким образом, задача сводится к задаче о частице, преодолевающей потенциальный барьер высотой $\Delta U$ из метастабильного минимума.

Далее необходимо некоторое уточнение терминологии. Экспериментально установлено, что при высоких температурах скорость перехода следует зако-ну Аррениуса $\Gamma \propto \exp(-\Delta U / T)$, в то время как при $T \to 0$ она демонстрирует температурную независимость $\Gamma \propto \exp(-B)$, что и позволяет говорить о чисто квантовом поведении [8, 9]. Формально, можно определить некоторую критическую температуру $T_0 = \dfrac{\Delta U}{B}$, в области которой происходит плавный переход от квантового к термическому режиму, а саму ситуацию считать фазовым переходом второго рода от квантового к классическому поведению [1]. Однако, по-видимому, здесь речь идет лишь о «фазовом переходе» между двумя теориями. Поведение же самой системы будет оставаться квантовым при любых температурах в независимости от того, какую из теорий мы предпочитаем для решения данной задачи. Данное обстоятельство актуально как раз в мезоскопических системах, для которых проявление квантовых эффектов при высоких температурах особенно заметно [12].

Тем не менее, в [1] сообщается, что при достижении плавно изменяющимся поперечным полем некоторого критического значения ($h_x = 0.25$) в системе (2) может наблюдаться скачкообразный переход от квантового к термоактивационному режиму, который авторы [1] называют фазовым переходом I рода.

Проанализируем подробнее обоснованность этого вывода. В [1] потенциал (3) раскладывается в окрестности точки $x = 0$, причем авторы в разложении ограничиваются членами четвертого порядка

$$V(x) = -2(S+1/2)^2 D h_x + (S+1/2)^2 D \left[ \left(h_x^2 - h_x\right)x^2 + \left(\frac{1}{3}h_x^2 - \frac{1}{12}h_x\right)x^4 \right] + O(x^6) \quad (4)$$

Действительно, при переходе через $h_x = 1/4$ член четвертого порядка в (4) меняет знак, что резко изменяет характер $V(x)$ - исчезает «двухъямность» (рис. 2). Фактически, в этом случае уже нельзя говорить и о бистабильности системы в целом, что вызывает, как раз, сомнения в адекватности (4) физической ситуации.

На самом деле весь эффект «фазового перехода» обусловлен необоснованным пренебрежением членами высших порядков в (4). Если разложение было бы остановлено, скажем, на членах не четвертого, а шестого порядка, то для критического $h_x$ получилось бы уже значение 0.063 и т.д. Сам же по себе потенциал (3) не имеет каких бы то ни было особенностей во всем диапазоне $0 < h_x < 1$ (см. рис. 1). В пределе $h_x \to 1$ высота барьера, разделяющего метастабильные минимумы стремится к нулю и при $h_x = 1$ барьер исчезает. Физически это вполне объяснимо: величина внешнего поля сравнивается с полем анизотропии. В противоположном предельном случае $h_x \to 0$ ширина потенциального барьера стремится к бесконечности.

Таким образом, за счет действия поперечного магнитного поля физические свойства молекулярных магнитов типа $Mn_{12}Ac$ не могут претерпевать скачкообразных изменений (этот вывод справедлив, во всяком случае, в той мере, в какой гамильтониан (1) адекватен описываемой физической системе).

Братский государственный технический университет

Список литературы к статье О.П. Мартыненко, В. В. Махро, И.Г. Махро

О «фазовых переходах» в магнитных мезоскопических системах